\documentclass[conference,a4paper]{APSIPA2020}
\usepackage{multirow}
\usepackage{graphicx}
\usepackage{amsmath}
\usepackage{subcaption}
\usepackage{verbatim}
\usepackage{cite}

\begin{document}

\title{Dynamic Noise Embedding: Noise Aware Training and Adaptation for Speech Enhancement}

\author{%
\authorblockN{%
Joohyung Lee, 
Youngmoon Jung, Myunghun Jung, 
Hoirin Kim
}
\authorblockA{%
School of Electrical Engineering, KAIST, Daejeon, South Korea \\
E-mail: {\{wngud701, dudans, kss2517, hoirkim\}}@kaist.ac.kr}
}

\maketitle
\thispagestyle{empty}

\begin{abstract}
Estimating noise information exactly is crucial for noise aware training in speech applications including speech enhancement (SE) which is our focus in this paper. To estimate noise-only frames, we employ voice activity detection (VAD) to detect non-speech frames by applying optimal threshold on speech posterior. Here, the non-speech frames can be regarded as noise-only frames in noisy signal. These estimated frames are used to extract noise embedding, named dynamic noise embedding (DNE), which is useful for an SE module to capture the characteristic of background noise. The DNE is extracted by a simple neural network, and the SE module with the DNE can be jointly trained to be adaptive to the environment. Experiments are conducted on TIMIT dataset for single-channel denoising task and U-Net is used as a backbone SE module. Experimental results show that the DNE plays an important role in the SE module by increasing the quality and the intelligibility of corrupted signal even if the noise is non-stationary and unseen in training. In addition, we demonstrate that the DNE can be flexibly applied to other neural network-based SE modules.
\end{abstract}

\section{Introduction}
Speech enhancement (SE) is a speech application which refines noisy speech into clean speech for improving the quality and the intelligibility of speech.
Conventional studies about SE have been based on statistical approaches such as spectral subtraction (SS) \cite{Boll1979}, Wiener-filtering \cite{Lim1979}, minimum mean-square error short-time spectral amplitude (MMSE-STSA) estimator \cite{Ephraim1984}, and subspace methods \cite{Dendrinos1991}. 

Recently, with the success of deep learning in various fields including speech applications, SE adopts deep learning-based methods and shows its potential. In time-frequency (T-F) domain, noisy magnitude spectrogram or log-power spectra (LPS) are extracted to be used as input acoustic features by applying short-time Fourier transform (STFT) to input noisy signal. Then corresponding enhanced features are derived by directly mapping the clean one or estimating the optimal T-F mask \cite{Lu2013, Wang2014, Xu2014, Xu2015, Weninger2015, Zhao2016, Park2017, Soni2018, Zhao2018, Fu2019}.

SE is used as a pre-processor in speech applications and improves the performance of main task by enhancing acoustic features. Therefore, many studies about combining SE with other speech application have been widely considered;
automatic speech recognition (ASR) \cite{Du2014, Ravanelli2016}, speaker verification (SV) \cite{Shon2019, Kataria2020, Jung2020}, voice activity detection (VAD) \cite{Wang2015, Jung2018}, etc.

Meanwhile, environmental characteristics such as type of noise or the degree of distortion, mainly expressed as signal-to-noise ratio (SNR), are main factors of degrading the performance in SE. Especially, noisy speech corrupted by non-stationary noise with low SNR is difficult to be recovered. Additionally, denoising the signal corrupted by unseen noise, which is not considered in training step, is also challenging issue. Therefore, to deal with these issues, researchers have used noise information in many speech applications, which is called noise aware training (NAT) \cite{Xu2014, Seltzer2013, Panahi2016, Qian2017}.

VAD is another pre-processor in speech applications. It is a framewise classification task which discriminates speech frames from non-speech frames. Results of VAD can be used in other speech applications to concentrate on speech frames \cite{McLaren2015, Jung2019} or implemented on devices without push-to-talk by being combined with post-processor like \emph{hangover scheme} \cite{Davis2006} for detecting utterance segments.

For these reasons, both SE and VAD are widely used as important pre-processors in speech applications. However, the order of using them, i.e., using SE first or VAD first, is always a \emph{chicken-and-egg problem}. The reason of using SE first is to enhance acoustic features for the input of the VAD module \cite{Wang2015, Jung2018} or to append hidden variables of SE module to input features of the VAD module \cite{Xu2019}. The case of using VAD first can be found in real-world devices; VAD is used to detect utterance segments in noisy signal and these segments are enhanced by the following SE module. In this case, because SE module is operated only on specific segments, computational costs are saved efficiently.

In this paper, we propose a deep learning-based novel method using both VAD and SE for masking-based single-channel denoising task. In the proposed method, VAD is used first to estimate the noise information and utilize it for NAT in SE module. In noisy speech, non-speech frames contain only noise component without speech, thus non-speech frames can be regarded as noise-only frames. Therefore, VAD can be used to detect non-speech frames. These estimated noise-only frames can provide the information about characteristics of noise. By using them with speech posteriors, simple neural network extracts the noise-adaptive embedding, which is called \emph{dynamic noise embedding} (DNE). The DNE is appended to input acoustic features of SE modules for improving the robustness in challenging noisy environment. 

The output of VAD has a great influence on the following SE module in our proposed method unlike conventional approaches where VAD and SE are operated independently. VAD and SE are jointly trained for optimization in the proposed method, thus, there is no need to pre-train the VAD or SE modules separately. Experimental results conducted on TIMIT dataset show that estimated noise-only frames by using VAD improve the performance of SE in noisy environments including unseen and non-stationary noise. Furthermore, using the proposed DNE as an auxiliary feature shows substantial improvement over previous approaches. In ablation study, we find the optimal threshold to detect noise-only frames. Moreover, various deep neural network-based SE modules improve their abilities by using the proposed DNE.

The rest of this paper is organized as follows. Section 2 introduces the proposed method to extract the DNE. Section 3 and Section 4 describe the SE and VAD module respectively. Section 5 represents the experimental setup and Section 6 shows the results of experiments. Then, Section 7 concludes the paper.

\section{Proposed Method}\label{proposed_method}
In this paper, since we focus on speech denoising task, only additive noise is considered and reverberation is not considered like close-talking application scenario. Therefore, noisy speech in T-F domain obtained by applying STFT to time-domain signal can be described as below.
\begin{gather}
    |Y_{t}|e^{j\phi^{Y}_{t}} = |X_{t}|e^{j\phi^{X}_{t}} + |N_{t}|e^{j\phi^{N}_{t}},\label{eq:basic}
\end{gather}
where $t$ denotes the frame index, and frequency bin index is omitted for brevity. $|Y_{t}|$, $|X_{t}|$, and $|N_{t}|$ are the magnitude spectrum of noisy speech, clean speech, and noise in the $t$-th frame, respectively. Likewise, $\phi^{Y}_{t}$, $\phi^{X}_{t}$, and $\phi^{N}_{t}$ are the phase spectrum of noisy speech, clean speech, and noise in the $t$-th frame, respectively. To make the equation more simple, it is assumed that the phase of the speech signal and noise signals are the same as did in approaches using ideal binary mask (IBM) \cite{Xu2014} and ideal ratio mask (IRM) \cite{Wang2014}. Then, (\ref{eq:basic}) is approximated as below to take into account only in magnitude. 
\begin{gather}
    |Y_{t}| \cong |X_{t}| + |N_{t}|.\label{eq:apprx}
\end{gather}
Because each frame is classified into 2 cases, speech frame or non-speech frame, (\ref{eq:apprx}) can be expressed as below.
\begin{gather}
    |Y_{t}|\cong\left\{\begin{array}{cc}\ |X_{t}| + |N_{t}| & \mbox{if\quad $t \in T_{S}$}\\|N_{t}| & \mbox{else}\end{array}\right.\,\label{noise_only_frames},
\end{gather}
where $T_{S}$ denotes the set of speech frames. In the case of ``else", the lower equation in (\ref{noise_only_frames}) means non-speech frames are considered as noise-only frames.

\begin{figure}[t]
\begin{subfigure}{.5\textwidth}
\centering{\includegraphics[clip, trim = {7.3cm 12.3cm 8.7cm 2.5cm}, width=\linewidth]{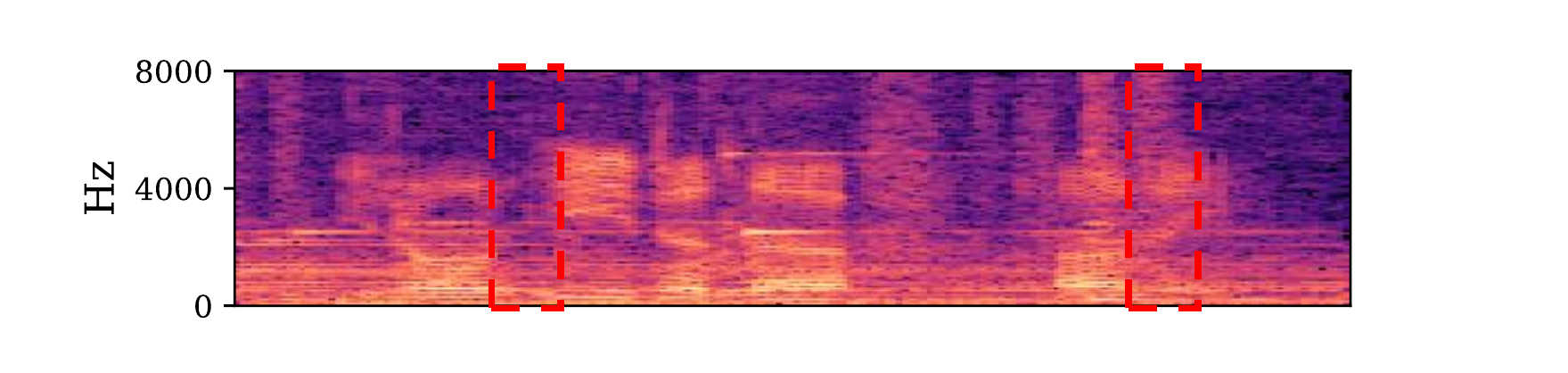}}
\vspace{-0.8cm}
\caption{Noisy spectrogram}
\label{noisy_mag}
\end{subfigure}
\begin{subfigure}{.5\textwidth}
\centering{\includegraphics[clip, trim = {7.3cm 8.4cm 8.7cm 6.5cm}, width=\linewidth]{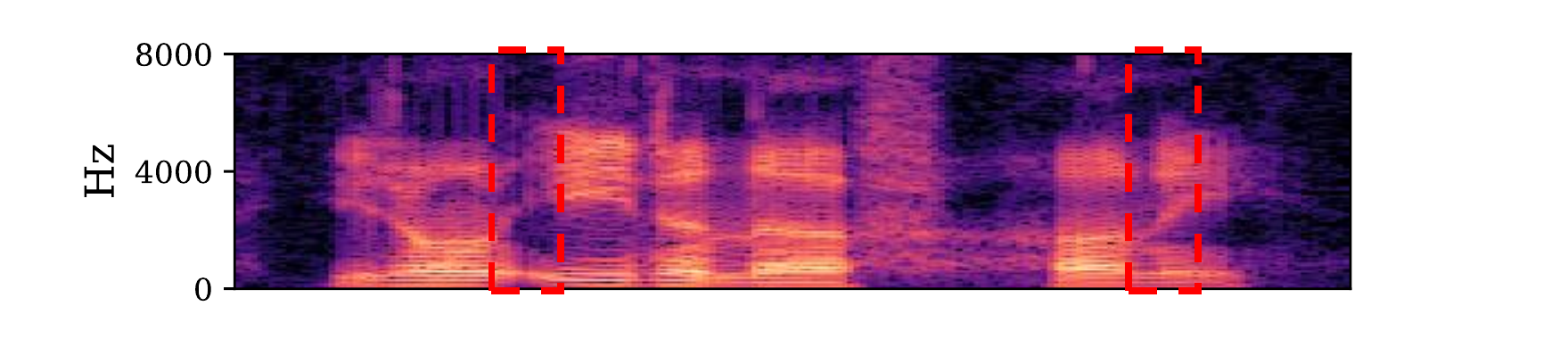}}
\vspace{-0.8cm}
\caption{Clean spectrogram}
\label{clean_mag}
\end{subfigure}
\begin{subfigure}{.5\textwidth}
\centering{\includegraphics[clip, trim = {7.3cm 5.2cm 8.7cm 10.1cm}, width=\linewidth]{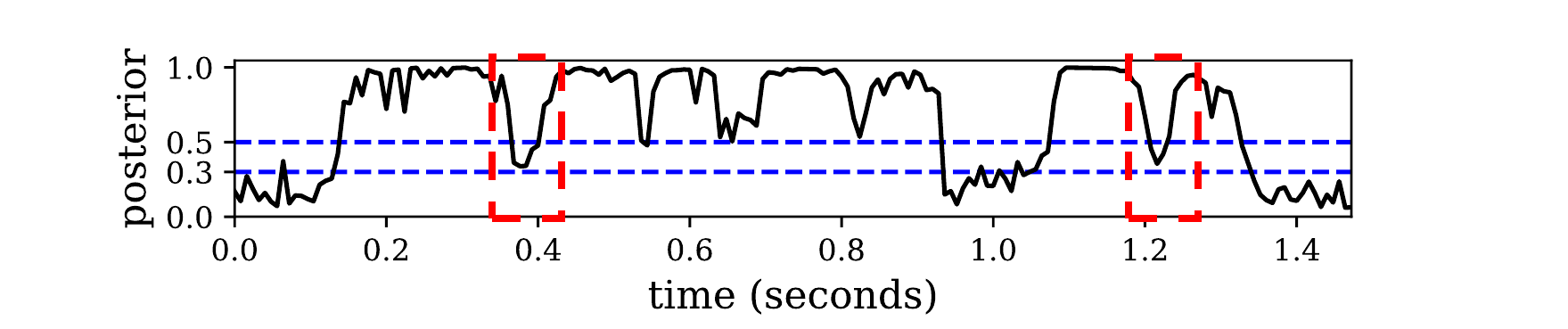}}
\vspace{-0.5cm}
\caption{Speech posterior}
\label{post.}
\end{subfigure}
\caption{Visualization of procedure to detect noise-only frames with 2 different thresholds. (a) and (b) represent the spectrogram of noisy and its original speech, respectively. (c) represents the speech posterior over time. In (c), the upper and lower blue dashed horizontal lines correspond to the thresholds of 0.5 and 0.3, respectively.}
\label{conf_noise}
\end{figure}

\begin{figure*}[t!]
    \centering{\includegraphics[clip, trim = {0cm 0.8cm 0cm 6.3cm}, width=\linewidth]{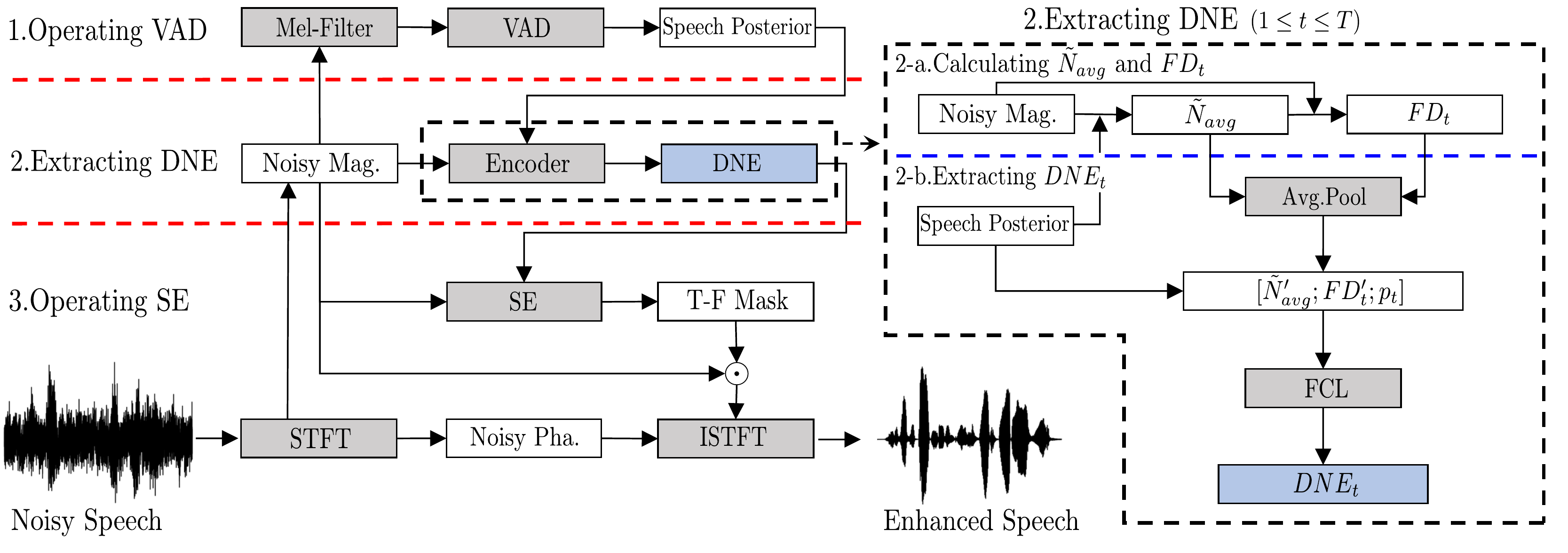}}
    \caption{The schematic diagram of the proposed method composed of 3 steps; operating VAD, extracting the DNE, and operating SE. The specific process of extracting the DNE is represented in right column enclosed with black dashed line. $\odot$ between T-F Mask block and ISTFT block denotes the element-wise multiplication.}
    \label{Diagram}
\end{figure*}

\subsection{Estimating Confident Noise Frames}
It is crucial for NAT to exactly speculate the noise information from input noisy utterance. In \cite{Xu2014, Seltzer2013, Qian2017}, the noise information is estimated by just averaging the several frames at the beginning and end of the utterance. This method is simple but it is hard to represent the tendency of non-stationary noise. In addition, those frames are not guaranteed that they are always noise-only frames. 

As we can see in (\ref{noise_only_frames}), non-speech frames can represent the noise information helpfully. Therefore, in this work, we propose to use Long Short-Term Memory (LSTM)-based VAD to estimate non-speech frames exactly. If we can detect non-speech frames as exactly as possible, we can use the noise information more precisely for NAT. 

To detect non-speech frames, we obtain the \emph{speech posterior}, which is the output of the VAD, at first. The mathematical expression of speech posterior from a VAD module can be represented as below.
\begin{gather}
    p_{t} = f_{\mathcal{VAD}}(g(|Y_{t}|))\quad for\quad1\leq t\leq T,
\end{gather}
where $p_{t}$ denotes the speech posterior of the $t$-th frame and $T$ denotes the total number of frames in the utterance. The hidden state of LSTM is omitted for brevity.
Function $g(\cdot)$ converts $|Y_{t}|$ to the input features of the VAD, such as mel-frequency cepstral coefficients (MFCCs) or mel-filter bank energies (MFBs). Function $f_{\mathcal{VAD}}(\cdot)$ is LSTM-based VAD function which takes acoustic features as input and estimates the speech posterior of each frame. 

After getting speech posteriors by operating a VAD module, we can choose non-speech frames by selecting frames whose posteriors are smaller than pre-defined threshold, $\eta$. If the threshold is set as 0.5, the median value of posterior, some speech frames can be misclassified as non-speech frames (\emph{false negative}). However, if the threshold value is low, only frames whose posteriors are fairly small are determined as non-speech frames. Therefore, in the experiment, we set threshold under 0.5. In this case, if speech posteriors of frames are smaller than the threshold value, they are assumed to correspond to reliably noise-only frames. These estimated frames are called \emph{confident noise frames}. This approach is motivated by \cite{Jung2019} which makes reliable speech / non-speech label for domain adaptation in VAD. 

Fig. \ref{conf_noise} shows the difference of selecting noise-only frames along different thresholds, 0.5 and 0.3, which corresponds to the upper and lower blue dashed horizontal lines in Fig. \ref{post.}, respectively. When threshold is set as 0.5, some speech frames are determined as noise-only frames. However, when threshold is set as 0.3, these frames are not determined as noise-only frames (indicated by red dashed rectangles).

\subsection{Dynamic Noise Embedding}
In \cite{Panahi2016}, VAD is used to detect noise frames by concentrating on the absence of speech frames, which is similar to our proposed method. Estimated noise frames are used to classify the type of noise, and an SE module is selected among 3 SE models (templates) depending on the classified noise type. However, in our proposed method, estimated noise frames are not used to classify the type of noise, but to capture the characteristic of noise. Then we extract the novel noise embedding, the DNE, and it provides the noise information to an SE module by using simple neural network. In this way, the SE module can be adapted to the environmental noise without using several SE models.

Confident noise frames can be used to feature vectors representing the characteristic of noise. These vectors are combined with a speech posterior and used to extract the DNE. The followings describe these feature vectors obtained by using confident noise frames and show the role of speech posterior in extracting the DNE.
\subsubsection{Confident Noise Average}
After detecting confident noise frames, these frames are averaged to represent noise information, which is called \emph{confident noise average}. It can be expressed as below.
\begin{gather}
    \tilde{N}_{avg} = {1 \over T'}\sum_{t=t_{n1}}^{t_{nT'}}|Y_{t}|\quad for\quad t_{n1}, t_{n2},...,t_{nT'} \in \tilde{T}_{N},\label{conf_noise_avg}
\end{gather}
where $\tilde{T}_{N}$ denotes the set of confident noise frames and $T'$ denotes the total number of confident noise frames. The confident noise average can represent the frequency characteristic of background noise.
\subsubsection{Framewise Difference}
If noise is stationary, confident noise average is sufficient to represent the characteristic of noise. On the other hand, if noise is non-stationary, it is hard to represent the characteristic of noise only with confident noise average. Therefore, we propose another vector which is helpful to represent information of non-stationary noise. It can be obtained simply as below.
\begin{gather}
    FD_{t} = ||Y_{t}| - \tilde{N}_{avg}|\quad for\quad 1\leq t\leq T,
\end{gather}
where $FD_{t}$ is \emph{framewise difference} of the $t$-th frame which means the difference between confident noise average and magnitude of the $t$-th frame. Framewise difference is obtained across whole frames in an utterance. The process of obtaining $\tilde{N}_{avg}$ and $FD_{t}$ is illustrated in 2-a of Fig. \ref{Diagram}.
\subsubsection{Speech Posterior}
Speech posterior shows the synergy effect when combined with confident noise average and framewise difference.
\begin{itemize}
\item In frames whose posteriors are relatively low; Low posterior means its frame is likely to be noise-only frame. Therefore, framewise difference shows steady or consistent tendency
in stationary noise, but it shows erratic tendency in non-stationary noise.
\item In frames whose posteriors are relatively high; High posterior means that its frame is likely to be speech frame, thus framewise difference has the meaning of segmental SNR (SSNR). It's because the greater the value of $|Y_{t}|$, the greater the value of framewise difference.
\end{itemize}
The DNE is extracted by simple neural network using confident noise average, framewise difference, and speech posterior. For reducing the dimension of noise average and framewise difference, average pooling is applied to down-sample them by half. These pooled values are represented as $\tilde{N}'_{avg}$ and $FD'_{t}$, respectively. Since we use 257-dim magnitude spectrogram, the dimension of $\tilde{N}_{avg}$ and $FD_{t}$ is 257 and that of $\tilde{N}'_{avg}$ and $FD'_{t}$ is 128. These reduced feature vectors are concatenated with a speech posterior and used to extract the DNE described as below. 
\begin{gather}
    DNE_{t} = f_{\mathcal{DNE}}([\tilde{N}'_{avg}; FD'_{t}; p_{t}])\quad for\quad 1\leq t\leq T,
\end{gather}
where $f_{\mathcal{DNE}}$ is a fully connected layer (FCL) for extracting the DNE. It has a single hidden layer with 128 hidden nodes and followed by leaky ReLU activation. The dimension of output node changes according to the backbone SE module, which will be explained in the next section, and activated by hyperbolic tangent function. The DNE is extracted for every frame and adaptive to environment with corresponding frame. Finally, the DNE is appended to input noisy magnitude spectrogram of an SE module for every frame. The process of extracting the DNE is illustrated in 2-b of Fig. \ref{Diagram}.

The proposed method is composed of 3 steps as illustrated in Fig. \ref{Diagram}. Firstly, a VAD module is operated by using noisy magnitude spectrogram (denoted as \emph{Noisy Mag.} in Fig. \ref{Diagram}) and speech posteriors are drawn. Secondly, the DNE is extracted by utilizing noisy magnitude spectrogram and posterior. At last, an SE module estimates the T-F mask by using the DNE as an auxiliary feature. Estimated mask is multiplied to noisy magnitude spectrogram and this enhanced magnitude spectrogram is combined with noisy phase spectrogram (denoted as \emph{Noisy Pha.} in Fig. \ref{Diagram}) for producing time-domain denoised signal by applying an inverse STFT (ISTFT). The specific procedure of extracting the DNE, which is in the second step, is illustrated in the right column of Fig. \ref{Diagram}. It is indicated by black dashed line. Configurations of SE and VAD modules are described in the following sections.
\section{Speech Enhancement Module}\label{SE}
For proving the effectiveness of the proposed method, we use 3 backbone SE modules based on deep neural network, mainly used in SE field. Noisy magnitude spectrogram standardized to have zero mean and an unit variance is used as input feature for all of SE modules. SE modules estimate the optimal T-F mask by activating final output with sigmoid function. Specific configuration and the way of utilizing the DNE in each model are described in the following sub-sections. 
\subsection{U-Net}
U-Net is a fully convolutional neural network (CNN) based on autoencoder with skip-connections. Although it was first introduced in medical image field \cite{Ronnenberger2015}, it has been shown its effectiveness in SE on T-F domain \cite{Ernst2018, Choi2019}. 
\begin{figure}[h!]
\centering{\includegraphics[clip, trim = {0.1cm 6.5cm 0.4cm 0cm}, width=\linewidth]{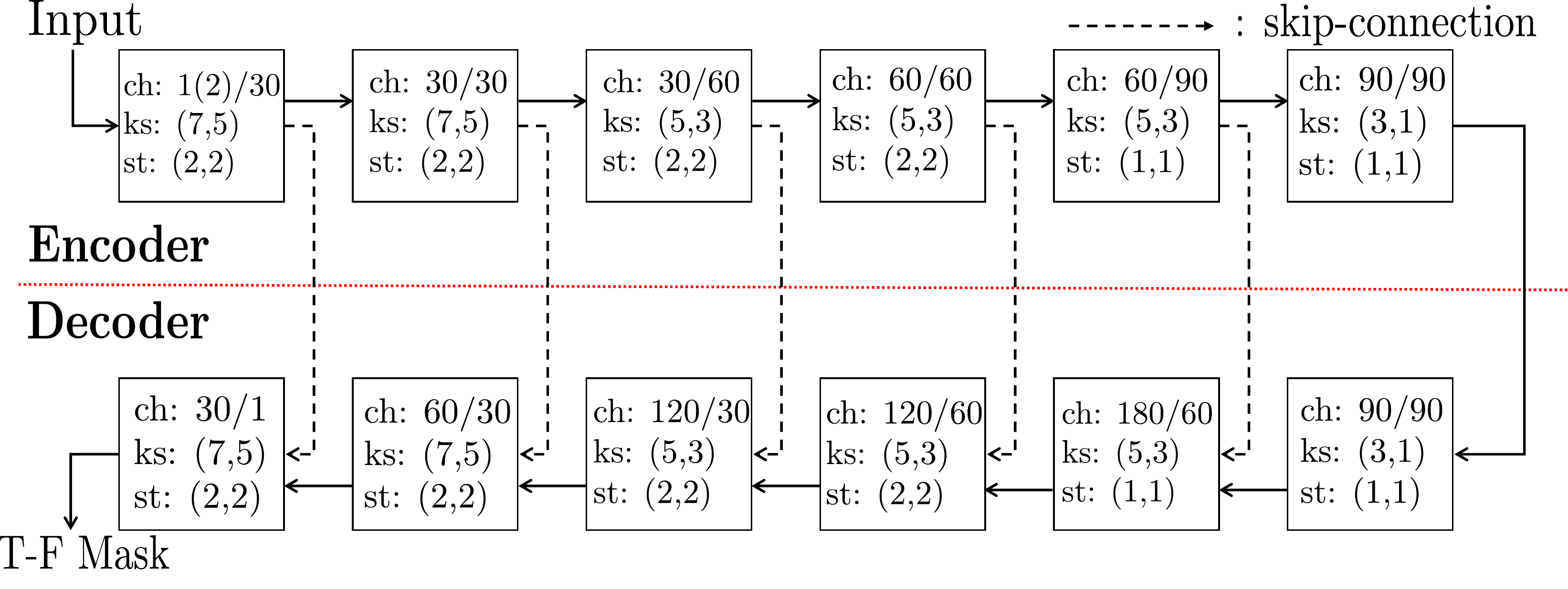}}
\caption{The configuration of U-Net. \emph{ch, ks, and st} denote the number of input / output channels, kernel size, and stride size, respectively.}
\label{unet_config}
\end{figure}

The configuration of U-Net used in this paper is described in Fig. \ref{unet_config}. Every convolution operation is followed by batch normalization and leaky ReLU activation except for last operation whose output is activated by sigmoid function.
\begin{table*}[t!]
\centering
\caption{Comparison of performances of baseline and various NAT-based models. In this paper, the best results are highlighted in bold. Averaged value is obtained in 2 groups and calculated over all the SNRs; seen noise (SEEN) and unseen noise (UNSEEN) environments.}
\begin{scriptsize}
\begin{tabular}{c|c|c||c|c|c|c|c|c|c|c|c|c}
\hline
\multicolumn{2}{c|}{\multirow{2}{*}{Noise type}} &
  \multirow{2}{*}{\begin{tabular}[c]{@{}c@{}}SNR\\ (dB)\end{tabular}} &
  \multicolumn{5}{c|}{PESQ} &
  \multicolumn{5}{c}{STOI} \\ \cline{4-13} 
\multicolumn{2}{c|}{}                                 &    & Noisy & U-Net & w. SN & w. CN          & w. DNE         & Noisy & U-Net & w. SN & w. CN & w. DNE         \\ \hline\hline
\multirow{10}{*}{SEEN}   & \multirow{3}{*}{Babble}     & -5 & 1.392 & 1.572 & 1.542 & 1.540          & \textbf{1.645} & 0.487 & 0.576 & 0.571 & 0.569 & \textbf{0.594} \\  
                         &                             & 0  & 1.507 & 1.933 & 1.914 & 1.938          & \textbf{2.003} & 0.613 & 0.735 & 0.728 & 0.734 & \textbf{0.747} \\ 
                         &                             & 5  & 1.779 & 2.366 & 2.335 & 2.380          & \textbf{2.437} & 0.726 & 0.836 & 0.833 & 0.839 & \textbf{0.844} \\ \cline{2-13} 
                         & \multirow{3}{*}{Factory1}   & -5 & 1.256 & 1.590 & 1.551 & 1.574          & \textbf{1.700} & 0.485 & 0.640 & 0.635 & 0.638 & \textbf{0.665} \\ 
                         &                             & 0  & 1.351 & 1.919 & 1.882 & 1.906          & \textbf{2.004} & 0.602 & 0.762 & 0.757 & 0.762 & \textbf{0.777} \\  
                         &                             & 5  & 1.521 & 2.274 & 2.251 & 2.295          & \textbf{2.342} & 0.722 & 0.847 & 0.846 & 0.849 & \textbf{0.855} \\ \cline{2-13} 
                         & \multirow{3}{*}{F16}        & -5 & 1.242 & 1.832 & 1.790 & 1.847          & \textbf{1.872} & 0.520 & 0.729 & 0.722 & 0.727 & \textbf{0.739} \\ 
                         &                             & 0  & 1.335 & 2.176 & 2.138 & 2.210          & \textbf{2.216} & 0.637 & 0.826 & 0.822 & 0.826 & \textbf{0.833} \\  
                         &                             & 5  & 1.523 & 2.525 & 2.499 & \textbf{2.577} & 2.551          & 0.756 & 0.889 & 0.887 & 0.891 & \textbf{0.894} \\ \cline{2-13} 
                         & \multicolumn{2}{c||}{Avg.}        & 1.434 & 2.021 & 1.989 & 2.030          & \textbf{2.086} & 0.616 & 0.760 & 0.756 & 0.759 & \textbf{0.772} \\ \hline\hline
\multirow{10}{*}{UNSEEN} & \multirow{3}{*}{Cafe}       & -5 & 1.280 & 1.423 & 1.391 & 1.416          & \textbf{1.585} & 0.490 & 0.578 & 0.580 & 0.581 & \textbf{0.608} \\  
                         &                             & 0  & 1.392 & 1.749 & 1.724 & 1.773          & \textbf{1.914} & 0.617 & 0.727 & 0.726 & 0.738 & \textbf{0.752} \\ 
                         &                             & 5  & 1.624 & 2.114 & 2.089 & 2.170          & \textbf{2.283} & 0.739 & 0.834 & 0.833 & 0.840 & \textbf{0.849} \\ \cline{2-13} 
                         & \multirow{3}{*}{Music}      & -5 & 1.526 & 1.798 & 1.781 & 1.823          & \textbf{1.912} & 0.625 & 0.698 & 0.690 & 0.697 & \textbf{0.703} \\ 
                         &                             & 0  & 1.722 & 2.125 & 2.094 & 2.131          & \textbf{2.239} & 0.705 & 0.781 & 0.777 & 0.783 & \textbf{0.793} \\ 
                         &                             & 5  & 2.044 & 2.558 & 2.533 & 2.584          & \textbf{2.638} & 0.793 & 0.860 & 0.857 & 0.863 & \textbf{0.868} \\ \cline{2-13} 
                         & \multirow{3}{*}{Machine gun} & -5 & 1.998 & 2.120 & 2.170 & 2.151          & \textbf{2.599} & 0.748 & 0.795 & 0.773 & 0.773 & \textbf{0.834} \\ 
                         &                             & 0  & 2.515 & 2.699 & 2.742 & 2.700          & \textbf{3.019} & 0.812 & 0.840 & 0.846 & 0.844 & \textbf{0.888} \\ 
                         &                             & 5  & 2.982 & 3.142 & 3.156 & 3.122          & \textbf{3.321} & 0.854 & 0.889 & 0.893 & 0.892 & \textbf{0.918} \\ \cline{2-13} 
                         & \multicolumn{2}{c||}{Avg.}        & 1.898 & 2.192 & 2.187 & 2.207          & \textbf{2.390} & 0.709 & 0.775 & 0.775 & 0.779 & \textbf{0.801} \\ \hline
\end{tabular}
\end{scriptsize}
\label{tab:exp1}
\end{table*}
In FCL or LSTM, because these networks use weight matrix that can fully capture input acoustic features, auxiliary features are appended along the axis of feature dimension. On the contrary, in CNN, since filters just focus on local information, the way of appending auxiliary feature should be different with FCL or LSTM. For using characteristic of CNN, the DNE is appended along the channel axis like as secondary channel of input features. This method is used in \cite{Hamid2014} as acoustic features with its first and second derivatives are concatenated to form 3-channel input feature maps for CNN based ASR model. If backbone is U-Net, the dimension of the DNE is set as dimension of noisy magnitude spectrogram, 257. Then the shape of the DNE is same with noisy magnitude spectrogram because the DNE is extracted as much as the number of total frame of noisy magnitude spectrogram. Hence, the first encoder unit in Fig. \ref{unet_config}, input channel is set as 1 for baseline U-Net and 2 for U-Net with the proposed DNE.
\subsection{Deep Denoising Autoencoder}
FCL based deep denoising autoencoder (DDAE) is also widely used in SE task \cite{Lu2013, Xu2014, Xu2015, Zhao2016}. In experiment, our DDAE takes 5 frames acoustic features as input to use contextual information and estimates the optimal T-F mask for central frame. For DDAE, the DNE is extracted with 128 dimension, the half of noisy magnitude spectrogram dimension. The DNE is appended to noisy magnitude spectrogram for every frame. Therefore, the input dimension without the DNE is 1,285 (257$\times$5) and the one with the DNE is 1,925 (385$\times$5). The hidden layers is consist of 7 layers with 1024, 512, 256, 128, 256, 512, and 1024 hidden nodes respectively and the dimension of output node is 257 for estimating the T-F mask. Batch normalization, ReLU function for activation, and drop-out with 0.2 probability are used at every hidden layers.
\subsection{Bidirectional LSTM}
Bidirectional LSTM (BLSTM) is also popular architecture for SE task \cite{Weninger2015, Fu2019}. BLSTM for our work has 2 hidden layers with 512 hidden nodes. For estimating the T-F mask, last hidden states with backward direction are selected and followed by FCL, composed of a single hidden layer with 300 hidden nodes, leaky ReLU function, and the output layer with 257 nodes for estimating the T-F mask of each time step. Like in DDAE, the DNE is extracted with 128 dimension and appended to noisy magnitude spectrogram for every frame. Therefore, input dimension in each time step is 257 without the DNE and 385 with the DNE.
\subsection{Optimization}
In all of SE modules, mean squared error (MSE) loss is used as criterion; Enhanced (masked) magnitude spectrogram is compared with its corresponding clean magnitude spectrogram. Also, Adam optimizer \cite{Kingma2015} with initial learning rate ($\alpha_{1}$) $10^{-3}$ is used and $\alpha_{1}$ is reduced by a factor of $10^{-1}$ with $10^{-8}$ of lower bound. The update for parameters of an SE module is expressed as below.
\begin{gather}
    \theta_{\mathcal{SE}} \leftarrow \theta_{\mathcal{SE}} - \alpha_{1}* l_{MSE},
\end{gather}
where $\theta_{\mathcal{SE}}$ and $l_{MSE}$ denote parameters and loss of an SE module, respectively. Additionally, $f_{\mathcal{DNE}}$, FCL for extracting the DNE, is optimized with an SE module.

\section{Voice Activity Detection Module}
\subsection{Configuration}
The VAD module is composed of unidirectional LSTM, 2 hidden layers with 64 hidden nodes. To use perceptual scale, we use 40-dimensional log MFBs as input acoustic features. Last hidden states of each time steps are followed by FCL, which is composed of a single hidden layer with 32 hidden nodes, ReLU function, and the output layer with single node. The final output is activated by sigmoid function and it represents the speech posterior of each time step.
\subsection{Optimization}
Cross entropy (CE) loss is used as criterion in the VAD module. The posterior of each frame is compared with its corresponding ground-truth. Optimizer for a VAD module is same with an SE module, Adam optimizer, but for initial learning rate ($\alpha_{2}$) $10^{-2}$. For jointly training both modules, the VAD module is influenced by MSE loss in an SE module as well as CE loss. The process of optimization for the VAD module is expressed as below.
\begin{gather}
    \theta_{\mathcal{VAD}} \leftarrow \theta_{\mathcal{VAD}} - \alpha_{2}*(l_{CE} + \lambda *l_{MSE}),
\end{gather}
where $\theta_{\mathcal{VAD}}$ and $l_{CE}$ denote parameters and loss of the VAD module. $\lambda$ is hyper-parameter for controlling the weight of $l_{MSE}$.

\section{Experimental Setup}
\subsection{Dataset}
TIMIT database \cite{Garofolo1988} is used for the experiments. It is composed of 4,620 utterances for training and 1,680 utterances for evaluation. To make noisy training set, we use 5 noise types (babble, factory1, F16, destroyer engine, and white) from NOISEX database \cite{Varga1993}. 1,250 utterances are randomly selected from training data and corrupted by those 5 noises with 4 SNR levels; -5, 0, 5, and 10 dB. In testing set, 3 seen noise types (babble, factory1, and F16) and 3 unseen noise types (cafe, music, and machine gun) are added to clean utterances. To show the robustness of our proposed method in various non-stationary noise types, we use cafe noise and music noise from other noise datasets, QUT noise (CAFE-CAFE-1) \cite{Dean2010} and MUSAN corpus (JAMENDO-3) \cite{Snyder2015}. Machine gun noise is from NOISEX database. 100 utterances are randomly selected from testing data and mixed with those 6 noises with 3 SNR levels; -5, 0, and 5 dB. As a result, training and testing set are composed of 25,000 utterances and 1,800 utterances, respectively. For seen noise, noise sources are splitted into 2 segments, the former one is used for training set and the latter one is used for testing set. The ground-truth of speech (1) / non-speech (0) label for noisy corpus is extracted by applying VQ-VAD \cite{vqvad2013} to its corresponding clean corpus.

\subsection{Setting}
All data are sampled at 16 kHz. STFT is calculated using Hann window with 32ms window length, 8ms hop length, and 512 FFT size. Thus the dimension of magnitude spectrogram is 257 as mentioned in Section \ref{proposed_method}. In the VAD module, to extract log MFBs, \emph{MelScale} function from torchaudio\footnote{https://pytorch.org/audio/transforms.html} library is applied to noisy magnitude spectrogram tensor. 
\subsection{Evaluation Metrics}
For evaluating the performance of the SE module, we use 2 metrics; perceptual evaluation of speech quality (PESQ) \cite{Rix2001} and short-time objective intelligibility (STOI) \cite{Taal2010}. These 2 metrics are widely used in SE to evaluate the quality and the intelligibility of enhanced speech, respectively. For both PESQ and STOI, the higher the better.

\section{Results}
\subsection{Effectiveness of the DNE in SE}
At first, we conduct an experiment to prove the effectiveness of the DNE in SE. U-Net is used as backbone architecture. To compare the proposed method with other approaches, we implement other 2 NAT-based models as well as our DNE-based model.
\begin{table}[t!]
\centering
\caption{Comparison of performances (PESQ / STOI) along the different thresholds for estimating confident noise section. $\eta$ denotes threshold.}
\begin{tabular}{c||c|c|c|c}
\hline
\multirow{2}{*}{$\eta$} & \multicolumn{4}{c}{SNR (dB)}                                 \\ \cline{2-5} 
                     & -5            & 0             & 5             & avg.          \\ \hline
0.2                  & 1.861 / 0.689 & 2.212 / 0.796 & 2.584 / \textbf{0.871} & 2.219 / 0.785 \\ \hline
0.3                  & \textbf{1.886} / \textbf{0.691} & \textbf{2.232} / \textbf{0.798} & 2.595 / \textbf{0.871} & \textbf{2.238} / \textbf{0.787} \\ \hline
0.4                  & 1.878 / 0.690 & \textbf{2.232} / 0.797 & 2.582 / 0.869 & 2.231 / 0.786 \\ \hline
0.5                  & 1.862 / 0.689 & 2.224 / 0.797 & \textbf{2.596} / \textbf{0.871} & 2.227 / 0.786 \\ \hline
1.0                  & 1.826 / 0.679 & 2.200 / 0.792 & 2.564 / 0.869 & 2.196 / 0.780 \\ \hline
\end{tabular}
\label{exp2}
\end{table}
\subsubsection{Simple noise (SN)}
As did in \cite{Xu2014, Seltzer2013, Qian2017}, we simply average the first 10 frames of each utterance. Averaged vector is broadcasted (copied) as noisy magnitude spectrogram for being concatenated as secondary channel feature map to noisy magnitude spectrogram. 
\subsubsection{Confident noise (CN)}
Confident noise average, $\tilde{N}_{avg}$ in (\ref{conf_noise_avg}), is used for auxiliary feature. As did in the SN, $\tilde{N}_{avg}$ is broadcasted and concatenated to noisy magnitude spectrogram.

For calculating the CN and DNE, we set threshold for speech posterior $\eta$ as 0.3 and weighting hyper-parameter $\lambda$ as 1.

Results are described in Tabel \ref{tab:exp1}. From this table, we can observe that the proposed DNE shows the best performance in all of situations except for PESQ in 5dB of F16 noise environment. Especially in unseen noise environments, the effect of the DNE is more remarkable. The relative improvements of PESQ and STOI compared to baseline (U-Net) are 2.62\% and 5.00\%, respectively in seen noise environment, however, 8.58\% and 11.6\%, respectively in unseen noise environment. 
It means the proposed method can analyze the environment even if the noise is mismatched with training step. Also, the performance of applying the DNE is sharply increased in machine gun noise compared to other environments. This means the DNE is appropriate for denoising the noisy speech corrupted by sporadic noise though it is also unseen noise type. 

Although both of the SN and CN use averaged value of estimated noise frames, the CN dominate the SN in most of situations. Besides, with focusing on the result of averaged value in both of seen and unseen environments, the CN improves the performance compared to baseline except for STOI in seen noise environments and improvements are more dominant in unseen noise environments like the DNE. However, the SN can't improve the performance. This means average of estimated noise frames can improve the performance of SE only if the section is reliable.

\subsection{Finding optimal threshold for the DNE}

As mentioned in Section \ref{proposed_method}, setting the optimal threshold under 0.5 is crucial for estimating confident noise frames. The best threshold is found by setting it variously, 0.2, 0.3, 0.4, 0.5, and 1.0, and comparing results. Threshold of 1.0 means whole frames are used to calculate noise average. That is to say, the VAD module is not used because it it not necessary. So, in this case, speech posteriors are set to value between 0 and 1 randomly. Table \ref{exp2} shows the result of experiment along the different threshold. At first, threshold of 1.0 shows the lowest results in all of situations. It represents that selecting specific frames for estimating noise section is more beneficial to NAT than just using whole frames. It can be found that setting threshold as 0.3 gets the best performances except for PESQ in 5dB which is just 0.001 difference with the best performance. Although threshold of 0.2 is more reliable than 0.3, its results are disappointing. It is because the number of selected frames is less in lower threshold and the number is insufficient to understand the background noise. On the contrary, in setting threshold as 0.4 or 0.5, their results are also unsatisfactory even if the number of selected frames is more than 0.3. It is because estimated noise frames are not reliable as described in Fig. \ref{conf_noise}. This is a kind of trade-off in setting threshold. 

\begin{figure}[]
\begin{minipage}[t]{0.48\linewidth}
    \includegraphics[clip, trim={2.25cm .5cm 16.25cm .5cm}, width=\linewidth]{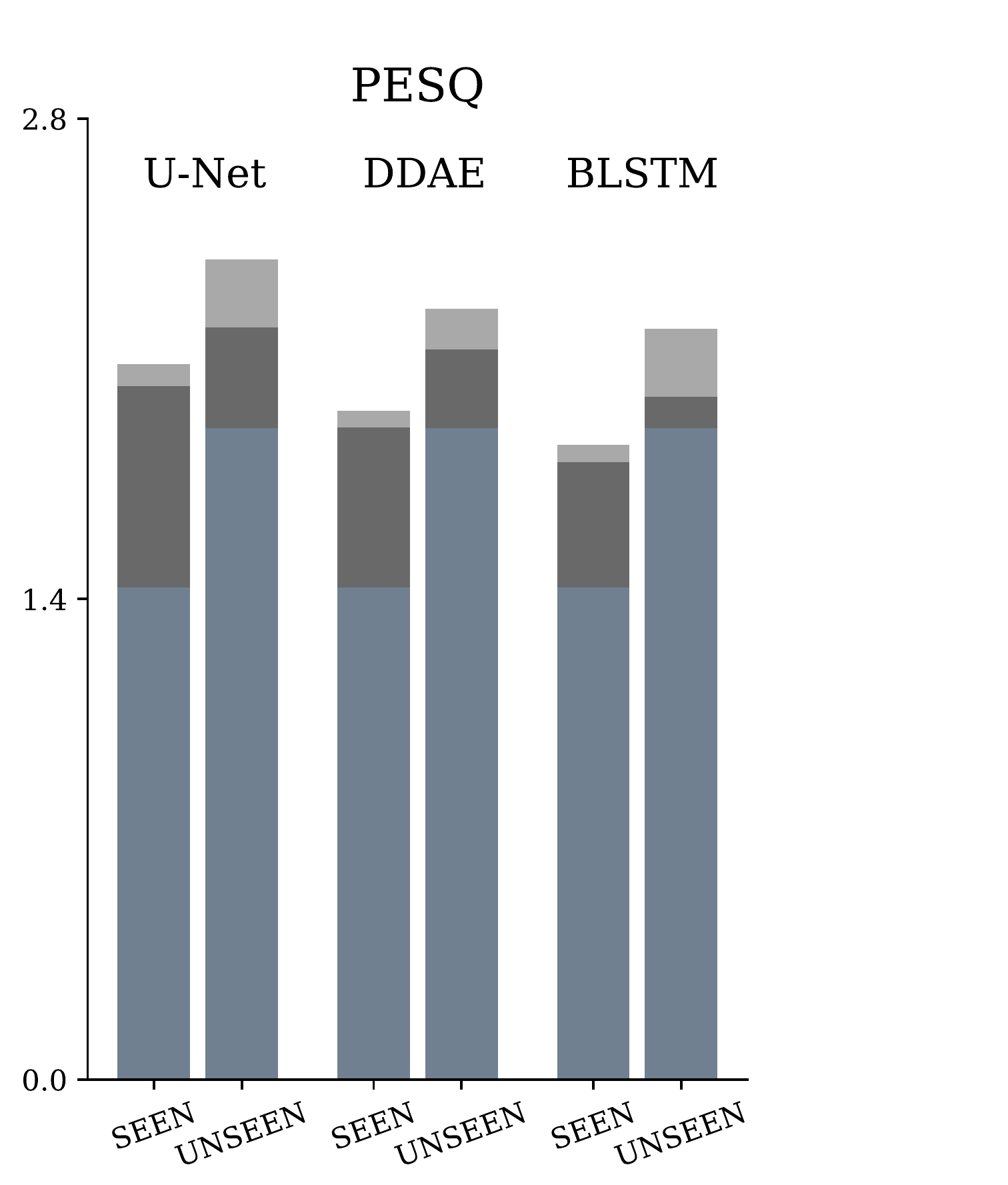}
    \label{f1}
\end{minipage}%
    \hfill%
\begin{minipage}[t]{0.48\linewidth}
    \includegraphics[clip, trim={2.25cm .5cm 16.25cm .5cm}, width=\linewidth]{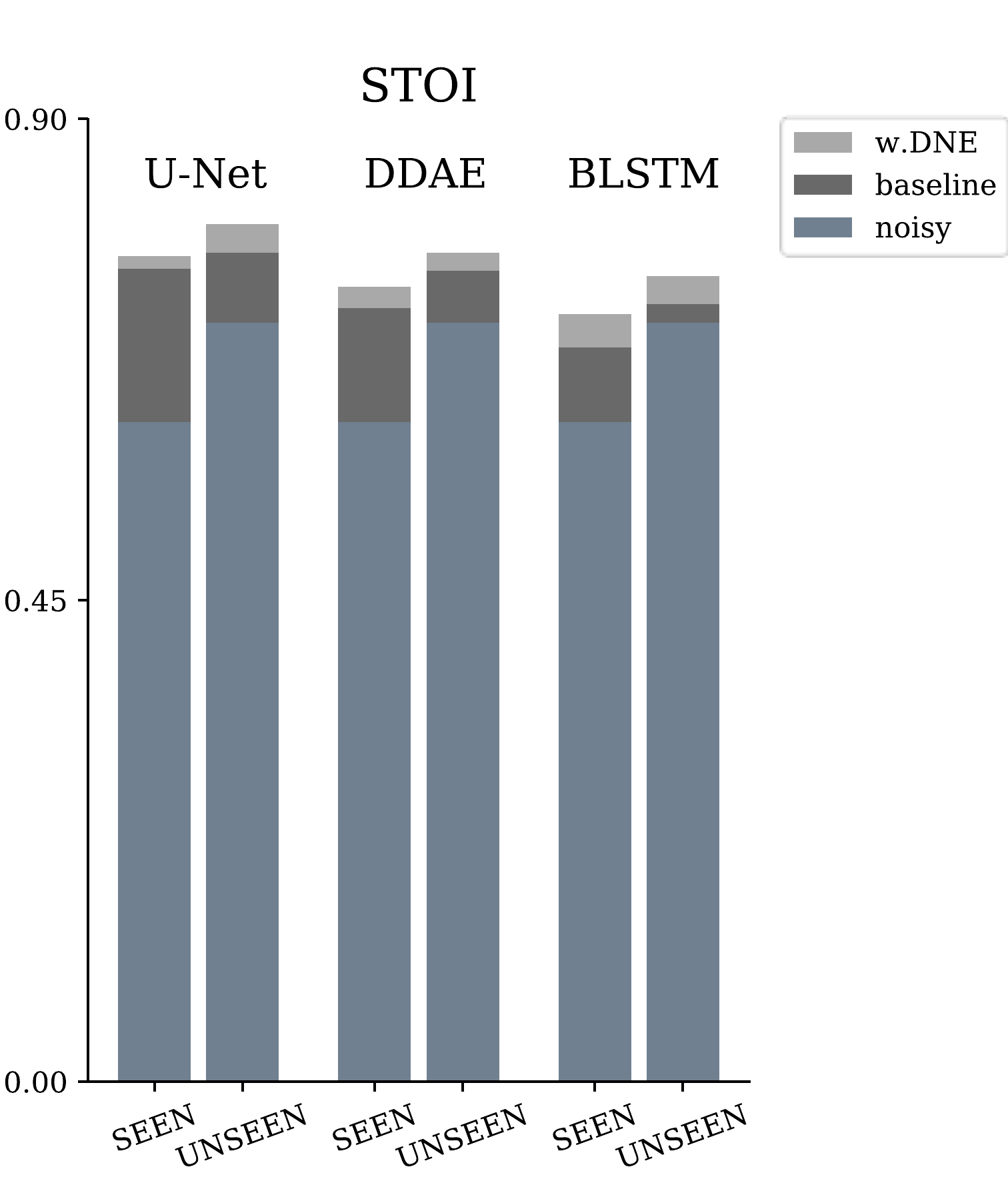}
    \label{f2}
\end{minipage} 
\caption{Performance of other neural networks in seen and unseen noise environments. The left one and right one represent PESQ and STOI, respectively.}
\label{bar_graph}
\end{figure}

\subsection{Expansion to other neural networks}
For proving the flexibility of the DNE, other neural networks are used as baseline. In this experiment not only U-Net but DDAE and BLSTM are used as backbone architecture. Configurations of each model and the way of appending the DNE are described in Section \ref{SE}. Threshold for speech posterior $\eta$ and weighting hyper-parameter $\lambda$ are set as 0.3 and 1, respectively. Fig. \ref{bar_graph} represents the results of the experiment. At first, we can observe that the DNE improves the performance in all of neural networks in both of seen and unseen noise environments. It can be said that it is beneficial to incorporate the DNE to all of FCL, CNN, and LSTM based SE modules. In PESQ, the performance is increased by a large margin in unseen environments than seen environments. In STOI, the increase of performance is similar in both of environments for DDAE and BLSTM, but in U-Net, increase of STOI is bigger in unseen noise environments like in PESQ.
\section{Conclusions}
In this paper, we proposed a novel SE method using the noise embedding named DNE. With the DNE, an SE module can be adapted to background noise to improve the noise robustness. Specifically, we used VAD to detect non-speech frames, thus obtaining noise information. After that, the DNE is extracted by using the noise information with simple FCL. Because the DNE is extracted by utilizing noise information and optimized with an SE module jointly, the SE module can be adapted to environmental noise. The proposed method achieved better performances than baseline and other approaches in the TIMIT database. Especially, this method showed the robustness in non-stationary and unseen noise environments. Furthermore, the DNE can be flexibly applied to various deep neural network-based SE modules. All SE modules performed better when using the DNE as an auxiliary feature. In the future, we will utilize the speaker embedding as well in SE to handle the speaker variations.
\section*{Acknowledgment}
This work was conducted by Center for Applied Research in Artificial Intelligence (CARAI) grant funded by DAPA and ADD (UD190031RD).


\end{document}